**Odd-parity magnetoresistance in pyrochlore iridate thin films with broken time-reversal symmetry**

T. C. Fujita[1], Y. Kozuka[1,*], M. Uchida[1], A. Tsukazaki[1,2,3], T. Arima[4,5], M. Kawasaki[1,5].

[1]**Department of Applied Physics and Quantum-Phase Electronics Center (QPEC), University of Tokyo, Tokyo 113-8656, Japan.**

[2]**Institute for Materials Research, Tohoku University, Sendai 980-8577, Japan.**

[3]**PRESTO, Japan Science and Technology Agency (JST), Tokyo 102-0075, Japan.**

[4]**Department of Advanced Materials Science, University of Tokyo, Kashiwa 277-8561, Japan.**

[5]**RIKEN Center for Emergent Matter Science (CEMS), Wako 351-0198, Japan.**

**A new class of materials termed topological insulators have been intensively investigated due to their unique Dirac surface state carrying dissipationless edge spin currents. Recently, it has been theoretically proposed that the three dimensional analogue of this type of band structure, the *Weyl Semimetal* phase, is materialized in pyrochlore oxides with strong spin-orbit coupling, accompanied by all-in-all-out spin ordering. Here, we report on the fabrication and magnetotransport of $Eu_2Ir_2O_7$ single crystalline thin films. We reveal that one of the two degenerate all-in-all-out domain structures, which are connected by time-reversal operation, can be selectively formed by the polarity of the cooling magnetic field. Once formed, the domain is robust against an oppositely polarised magnetic field, as evidenced by an unusual odd field dependent term in the magnetoresistance and an anomalous term**

**in the Hall resistance. Our findings pave the way for exploring the predicted novel quantum transport phenomenon at the surfaces/interfaces or magnetic domain walls of pyrochlore iridates.**

A zero-gap surface state protected by time-reversal symmetry is formed at the surface of topological insulators, where electrons with up or down spins counter-propagate[1-6]. When a topological insulator is doped with magnetic elements[5-7], these surface states may become gapped, resulting in the dissipationless flow of a single chiral edge mode which effectively exhibits the $\nu = 1$ quantum Hall state in the absence of an external magnetic field[8,9]. A similar zero-gap state in three dimensional momentum space, referred to as Weyl Semimetal (WSM) phase[10-13], has recently been predicted to emerge in a certain class of pyrochlore oxide with broken time-reversal symmetry[14,15]. As WSM possesses a topologically non-trivial band structures provided momentum is conserved, a unique surface state accompanied with edge currents is predicted at interfaces with trivial insulators as in the case of topological insulators[15-17].

Pyrochlore oxides, formulated as $A_2B_2O_7$, are comprised of cation sublattices forming a corner-sharing tetrahedron network (Fig. 1a; ref.18) where antiferromagnetically coupled spins at the cation sites cannot determine their configurations uniquely due to geometrical frustration. In the case of lanthanides (*Ln*) iridate pyrochlore ($Ln_2Ir_2O_7$; $A = Ln$, $B = $ Ir), Dzyaloshinski-Moriya interaction enhanced by strong spin-orbit coupling can stabilize an all-in-all-out antiferromagnetic spin order for each cation sublattice[15,19,20], in which tetrahedra with all four spins at the vertices pointing inward or outward are alternatingly arranged along <111> direction, as shown in Fig. 1b. A notable feature of this spin structure is while characterised as antiferromagnetic, two distinct spin arrangements exist, of which are interchangeable by time-reversal operation. We hereafter refer to these two spin arrangements as A-domain and B-domain. This characteristic antiferromagnetic ordering has begun to attract attention as a key ingredient to substantiate WSM and to explore the surface states at the boundaries between A- and B-domains[21].

To investigate such intriguing phenomena through electrical measurements, it is necessary to control the magnetic domains and the carrier density of a single crystal. A promising approach to realising this is the growth of thin films, as it may enable the control of artificial magnetic domain boundaries and the exploration of carrier transport at such interfaces, as well as tunability of the Fermi energy at Weyl point by pursuing field-effect transistor devices. Nevertheless, obtaining high-quality $Ln_2$Ir$_2$O$_7$ single crystal has remained challenging even in bulk because of the low reactivity of metallic iridium and the volatility of iridium peroxides[22]. Here, we report the fabrication of single crystal pyrochlore iridate thin films, Eu$_2$Ir$_2$O$_7$ ($Ln$ = Eu), and their magnetotransport properties mediated by all-in-all-out spin ordering. We chose $Ln$ = Eu because of the total magnetic moment $J = 0$ of Eu$^{3+}$ for among $Ln_2$Ir$_2$O$_7$ series, this compound is a sole and ideal platform for studying carrier transport in the background of all-in-all-out spin structure composed of the Ir$^{4+}$ moment in $J_{\text{eff}} = 1/2$ state[23]. The results indicate that the magnetic domain structure is detected via peculiar asymmetric term in the magnetoresistance (MR) and zero-field offset in Hall resistance. These observations provide controllability of the exotic electronic phase in this compound towards accessing WSM.

**Results**

**Sample fabrication and structural property**.

In this study (111)-oriented $Eu_2Ir_2O_7$ thin films were epitaxially grown on Y-stabilized $ZrO_2$ (YSZ) (111) single crystal substrates by pulsed laser deposition (see Methods and Supplementary Fig. S1). Transmission electron microscopy (Figs. 1c and 1d) confirms the formation of a single-crystal $Eu_2Ir_2O_7$ film free from crystalline domain boundaries. The macroscopic crystal structure was examined by X-ray diffraction (XRD), with the high quality of the films supported by the observation of a typical rocking curve of 0.09° for the $Eu_2Ir_2O_7$ (222) peak (Fig. 1e and Supplementary Fig. S2). In Fig. 1f, the reciprocal space mapping is shown, indicating that the lattice of the film is elongated along [111] direction by 0.7 % with respect to in-plane lattices. The surface morphology was measured by atomic force microscope, which showed root mean square roughness of ~ 1 nm before and after post-growth annealing (Supplementary Fig S3). The film thickness is fixed at about 70 nm.

**Longitudinal resistivity.**

Figure 2a displays the temperature dependence of the longitudinal resistivity ($\rho_{xx}$). For electrical measurements, the sample was defined into a Hall-bar geometry (inset of Fig. 2b) to reduce mixing of the longitudinal and Hall resistances. A metal-insulator transition (MIT) is observed around transition temperature ($T_M$) of 105 K, which is close to the reported value for bulk (120 K, ref. 24). The absolute values of $\rho_{xx}$ and the strength of the MIT sensitively depend on the growth conditions (data for typical films are shown in Fig. S4a), likely as a result of film-dependent Eu/Ir nonstoichiometry. Since analytical determination on the composition of thin films is challenging, we estimate the composition of films by comparing the resistivity ratio $\rho_{xx}(2\ K)/\rho_{xx}(300\ K)$ with those of previously reported polycrystalline bulk data[25]. The result indicates that the films are Ir-rich by 1– 4 % (Fig. S4b). The reduction of $T_\mathrm{M}$ compared with the bulk may also be explained by the cation nonstoichiometry. Irrespective of quantitative variations in $\rho_{xx}$, we qualitatively obtain the same magnetotransport properties discussed later for all the thin films. Hence, we focus on the most conducting film (sample 3 in Supplementary Fig.S4) in the structural and transport data presented. As also noted in Ref. 25 for polycrystalline bulk samples, the temperature dependence of Hall coefficient ($R_\mathrm{H}$) estimated by Hall measurement (see Methods) does not show significant anomaly across $T_\mathrm{M}$, in contrast to the sharp kink at MIT in $\rho_{xx}$ as shown in Fig. 2b. Although the origin of this behaviour remains controversial[26,27], we simply refer to this transition as the MIT.

**Magnetoresistance.**

Magnetotransport experiments provide versatile information reflecting the orbital motion and spin states of electrons, as exemplified by the spin liquid recently detected by the zero-field anomalous Hall effect in $Pr_2Ir_2O_7$ (ref. 28). Figure 3a shows the out-of-plane MR at 2 K. The MR for zero-field cooling (left panel) shows a symmetric shape with respect to zero field with a roughly 0.5% decrease in resistance at high magnetic field - a typical response in magnetic materials. The sample cooled with an applied magnetic field, however, unexpectedly exhibits an asymmetric term in MR in addition to the symmetric term (right panel). Although the symmetric term is found to be independent of cooling field (Fig. S5), the sign of the asymmetric term is changed by inverting the polarity of the cooling field. We emphasize that this asymmetric term is an intrinsic property of $\rho_{xx}$ and is not caused by intermixing from Hall resistivity ($\rho_{yx}$), as evidenced by the fact that the two-terminal MR shows qualitatively the same asymmetric term (Supplementary Fig. S6).

This observation appears peculiar as MR usually does not include an asymmetric term and is not dependent on the cooling field. Although magnetic metals or semiconductors show field asymmetric MR around zero field with a hysteretic behaviour due to a finite coercive field, it is symmetric when spins are flipped at high magnetic field. The MR observed in the data presented here can phenomenologically be understood by taking into account the all-in-all-out spin structure, based on the double exchange model as in the case of perovskite manganites. Here, the angle between the neighbouring spins dramatically affects the hopping probability, giving rise to MR[29]. For example, if A-domain is stabilized with cooling under +9 T, the spins on basal planes are canted towards the in-plane direction with positive magnetic field resulting in an increase in $\rho_{xx}$ (top right inset). Conversely, a negative magnetic field tends to align those spins, giving rise to a lower $\rho_{xx}$ (bottom left inset). The opposite response arises in the case of B-domain with $-9$ T field cooling.

**Temperature and cooling field dependence of the linear component in $\rho_{xx}$.**

To further explore this idea, the same measurements are systematically performed as a function of temperature and cooling field (see Methods). Here, we define a parameter $\alpha$ to describe the sign and the magnitude of the asymmetric term as $[\rho_{xx}(B) - \rho_{xx}(0)] / \rho_{xx}(0) =$ (even term) $+ \alpha B$, since the odd terms equal to or higher than the third order are not significant (less than 1 % of the linear term by fitting at 2 K). Figure 3b shows $\alpha$ as a function of cooling field at 2 K, indicating that $\alpha$ saturates above the cooling field of $\pm 3$ T. This bistability suggests that A-domain or B-domain may be selectively stabilized by applying a cooling field above ±3 T. The temperature dependence of $\alpha$ shown in Fig. 3c is also consistent with the above scenario as $\alpha$ disappears above $T_M$. Although small hysteresis is observed around 60 K, the asymmetric term is clearly observed even in such a temperature region (Supplementary Fig. S7), indicating that one magnetic domain is dominant over the other.

**Hall measurements.**

So far we have demonstrated that the MR contains a linear term reflecting the formation of a domain structure depending on the cooling field. We now show that the Hall resistance additionally captures the lattice distortion through the anomalous terms. $\rho_{yx}$ is generally expressed as

$$\rho_{yx} = R_H B + R_A M + \rho^T, \qquad (1)$$

with $R_H$ the ordinary Hall coefficient, $R_A$ the Hall coefficient proportional to the macroscopic magnetization $M$ as a result of spin-orbit interaction, and $\rho^T$ the topological term induced by spin chirality $\chi = \mathbf{S}_i \cdot (\mathbf{S}_j \times \mathbf{S}_k)$ (ref. 30). Finite $M$ and $\chi$ may remain in the presence of uniaxial lattice distortion induced by epitaxial strain, while they should be cancelled out in the case of ideal all-in-all-out spin structure with cubic symmetry. Figure 4a shows the raw data of the Hall measurement at 2 K for the cooling magnetic fields of ±9 T. In addition to the ordinary $B$-linear component common in both measurements, the Hall resistance clearly exhibits a non-zero value at 0 T, as measured by the difference in $\rho_{yx}$ between the opposite cooling magnetic fields ($\Delta\rho_{yx}$). The overall linear term is reasonably assigned to the ordinary

Hall effect (the first term in Eq. (1)), since no anomaly is present across $T_M$ as shown in Fig. 2b. We can therefore ascribe $\Delta\rho_{yx}$ to the sum of the second and third terms in Eq. (1) as $\Delta\rho_{yx}$ disappears above $T_M$ (Fig. 4b).

**Discussion.**

As opposed to the standard field-symmetric MR typically observed in prevalent conductors, the magnetoresponse in the presence of all-in-all-out spin ordering is described in terms of a third rank tensor. Therefore, odd terms should characteristically be observed in response functions due to time-reversal symmetry breaking as proposed by Arima[31], which is a clear fingerprint of the all-in-all-out spin ordering. Here, conduction electrons must be sensitive to the local spin structure of the compound owing to spin-orbit coupling, known as the Kondo lattice model. Hence, our observation substantiates that MR reflects not only the magnetic s*pin* structure but also the magnetic *domain* structure. From these considerations, the magnetic domain structure in $Eu_2Ir_2O_7$ is robust against a magnetic field of at least 9 T below $T_M$. This is in marked contrast to the case of $Nd_2Ir_2O_7$ where A-domain and B-domain are thought to be switchable through the Ising character of Nd ions by sweeping magnetic field[32]. Therefore, it is plausible to attribute the robust domain structure to the $J = 0$ character of $Eu^{3+}$.

The anomalies found in $\rho_{yx}$ are consistent with the interpretation of MR data in which conducting carrier reflects the magnetic domain structure in another way, and additionally indicate that the lattice distortion actually induces macroscopic $M$ and $\chi$. In addition to $\Delta\rho_{yx}$, there exists a low-field nonlinear component ($\rho_{NL}$), which is deduced by subtracting the linear term and the zero-field offset ($\Delta\rho_{yx}$) as shown in

Fig. S8. Remarkably, $\rho_{NL}$ shows sign inversion between the two domains. If we assume that $M$ is proportional to $B$ as in the case of the bulk[25], $\rho_{NL}$ may be therefore attributed to $\rho^T$ possibly originating from spins at the surface or the interface. Although this sign inversion in $\rho_{NL}$ is an intriguing character of this system, the details remain to be investigated including the possibility of band crossing as in the case of $EuTiO_3$ (ref. 33) or $SrRuO_3$ (ref, 34).

In conclusion, we have fabricated single crystal $Eu_2Ir_2O_7$ thin films via solid phase epitaxy and systematically explored the magnetotransport properties originating from the all-in-all-out spin order. The MR contains a linear field dependent term while the Hall resistance has zero field offset and anomalous bending, both of which inherently depend on the type of domain formed. Our findings indicate that the all-in-all-out spin structure is controllable by the polarity of the cooling field and once formed is robust below $T_M$. The ability to realize this spin structure in thin film form will become the cornerstone to realizing WSM phases via tailored pyrochlore iridate heterostructures and/or utilizing electrostatic gating.

**Methods**

The $Eu_2Ir_2O_7$ (111) films were prepared on commercial YSZ (111) single crystal substrates by pulsed laser deposition. Phase-mixed ceramic targets fabricated by hot-press method at 950 °C under 25 MPa pressure were used with a prescribed ratio of Eu/Ir = 1/3. The films were deposited at a substrate temperature of 500 °C in an atmosphere of 100 mTorr Ar gas containing 1 % $O_2$. A KrF eximer laser ($\lambda$ =248 nm) was used for ablating the target, with a fluence and frequency of 6 J/cm$^2$ and 10 Hz. The films were in an amorphous phase after the deposition, with the pure

Eu$_2$Ir$_2$O$_7$ phase appearing after annealing in an electrical muffle furnace at 1000 °C for 1.5 hours in air.

The Hall bar structure was processed by conventional photolithography and Ar ion milling. Ohmic Ni (10 nm) / Au (50 nm) contacts were deposited by an electron-beam evaporator. Magnetotransport measurements were performed in a liquid He cryostat equipped with a 9 T superconducting magnet (PPMS, Quantum Design Co.). The ordinary Hall coefficient ($R_H$) is calculated by linear term in Hall resistance above ± 5 T region. Temperature dependence measurements for $\alpha$ (Fig. 3c) were carried out by the following sequence. Magnetic field was applied at 150 K along [111] direction of YSZ substrate. Then the samples were cooled from 150 K down to 2 K under the fixed magnetic field. At 2 K, magnetotransport was measured as follows. In the case of positive (negative) field cooling, magnetic field was first set to −9 T (+9 T), then returned to +9 T (−9 T), and finally back to the initial cooling field. For example, in the case of +0.5 T cooling, magnetic field was swept as +0.5 T → −9 T → +9 T → +0.5 T. After this measurement, the sample was warmed to the next measurement temperature with +0.5 T of magnetic field.

**Acknowledgments**

We thank J. Fujioka, K. Ueda, Y. Tokura, N. Nagaosa, and Y. Motome for helpful discussions. We thank J. Falson for editing of the manuscript. This work was partly supported by Grant-in-Aids for Scientific Research (S) No. 24226002 and No. 24224010, for JSPS Fellows No. 26·10112 (TCF), for Challenging Exploratory Research No. 26610098 (MU) from MEXT, Japan, and by "Funding Program for World-Leading Innovative R&D on Science and Technology (FIRST)" Program from the Japan Society for the Promotion of Science (JSPS) initiated by the Council for Science and Technology Policy

**Author contributions**

T.C.F. performed the sample fabrication, measurements, and analysis. Y.K. assisted with the sample fabrication, measurements, and analysis. M.U., A.T. and T.A assisted with the planning and analysis. M.K. directed the project.
.

**Additional information**

Supplementary information accompanies this paper on http://www.nature.com/Scientificreports. Reprints and permissions information is available online at http://www.nature.com/reprints. Correspondence and requests for materials should be addressed to Y.K.

**Competing financial interests**

The authors declare no competing financial interests.

**Figure captions**

**Figure 1| Crystal and spin structures of the pyrochlore lattice. a**, The Ir sublattice of $Eu_2Ir_2O_7$. Kagome and triangular lattices formed by Ir are located at orange and yellow planes, respectively. **b**, Two distinct all-in-all-out spin structures, named as A-domain and B-domain, in the pyrochlore lattice. When four spins at tetrahedral vertices are consolidated at the centre of cube, they represent a magnetic octupole indicated by blue and red spheres. **c**, Phase contrast image of high-resolution TEM of a $Eu_2Ir_2O_7$ (111) film on a YSZ (111) substrate. **d**, Atomically resolved HAADF-STEM image at the $Eu_2Ir_2O_7$/YSZ interface. Triangular cross-sectional lattices composed of Ir and Zr are schematically shown. **e**, $\theta$-$2\theta$ scan of X-ray diffraction. Peaks from the substrate are marked with asterisks. **f**, The reciprocal space mapping around the YSZ (331) peak. The peak position of bulk $Eu_2Ir_2O_7$ is indicated by an open triangle. The solid line indicates that $Eu_2Ir_2O_7$ is coherently grown on the substrate and expanded along [111] direction by 0.7 %, which is illustrated by the pair of tetrahedra. An ideal pyrochlore lattice with cubic symmetry obeys the theoretical curve indicated by the dashed line.

**Figure 2 | Temperature dependence of the longitudinal resistivity $\rho_{xx}$ and Hall coefficient $R_H$. a**, Temperature dependence of the longitudinal resistivity $\rho_{xx}$ (left axis) and its temperature derivative (right axis). Magnetic transition temperature $T_M$ is estimated at the minimum of the resistivity and 105 K. **b**, Temperature dependence of the Hall coefficient $R_H$. The inset shows an optical microscope image of the sample and the transport measurements configuration.

**Figure 3 | Linear magnetoresistance mediated by all-in-all-out spin structure. a**, Magnetic field dependence of longitudinal resistivity $\rho_{xx}$ with $B$ // [111] at 2 K after

zero field cooling (left panel) and ±9 T cooling (right panel). While the zero-field cooled data shows a symmetric response, $\rho_{xx}$ after field cooling includes the linear terms indicated by dashed lines. Magnetic responses of $Ir^{4+}$ magnetic moments corresponding to respective magnetic domain structures are schematically shown as insets, where spins are depicted as solid arrows and magnetic domains are symbolized by the pair of open and shaded triangles. The initial points and the direction of the magnetic field sweeps are indicated by open circles and open arrows, respectively, for each measurement. **b**, Cooling field dependence of the linear magnetoresistance coefficient $\alpha$ at 2 K. $\alpha$ saturates above the cooling field of ±3 T, suggesting that magnetic domain be uniformly aligned to A-domain or B-domain. A multi-domain structure is formed by cooling in smaller magnetic fields. **c**, Temperature dependence of $\alpha$ for representative cooling magnetic fields, showing that $\alpha$ appears only below the magnetic transition temperature $T_M$.

**Figure 4 | Anomalies in Hall resistivity $\rho_{yx}$ originating from all-in-all-out spin structure. a**, Magnified view of magnetic field dependence of the Hall resistivity $\rho_{yx}$ with $B$ // [111] at 2 K after ±9 T field cooling. $\Delta\rho_{yx}$ is defined as the difference of $\rho_{yx}$ (0 T) between +9 T and -9 T field cooling. The scalar spin chirality induced by lattice distortion is schematically shown. The inset shows the data for the entire sweep range between ±9 T. **b**, Temperature dependence of $\Delta\rho_{yx}$. The transition temperature $T_M$ is indicted by a dashed line.

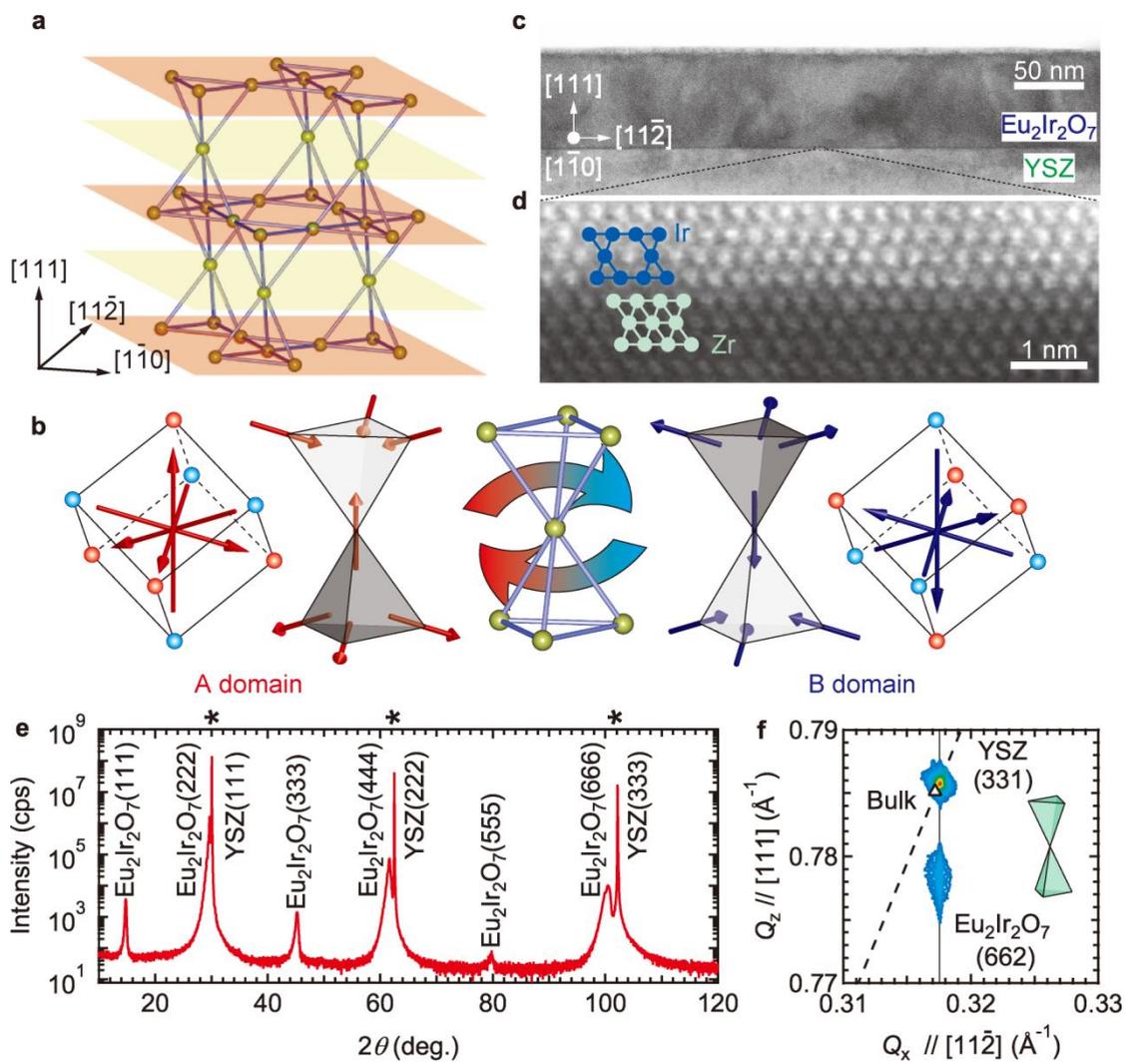

Fig. 1 Fujita et al.,

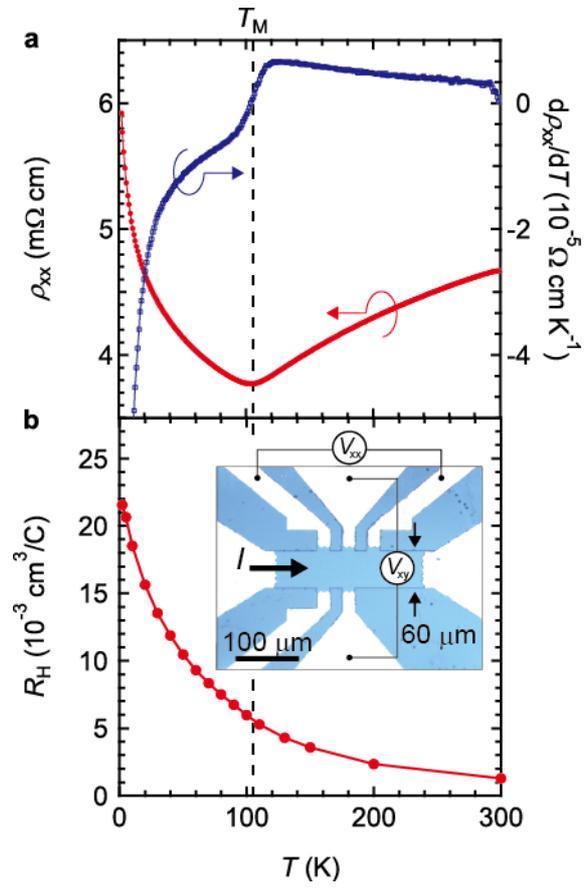

Fig. 2 Fujita et al.,

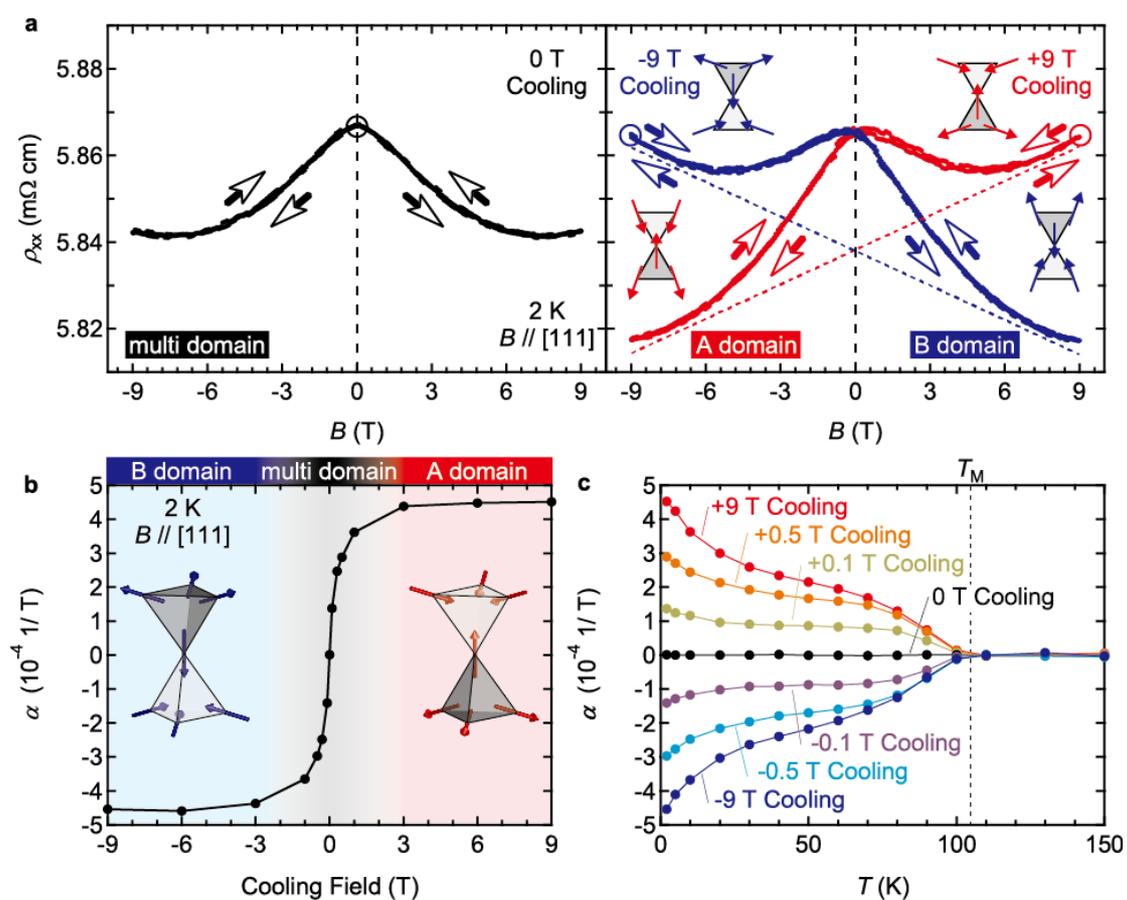

Fig. 3 Fujita et al.,

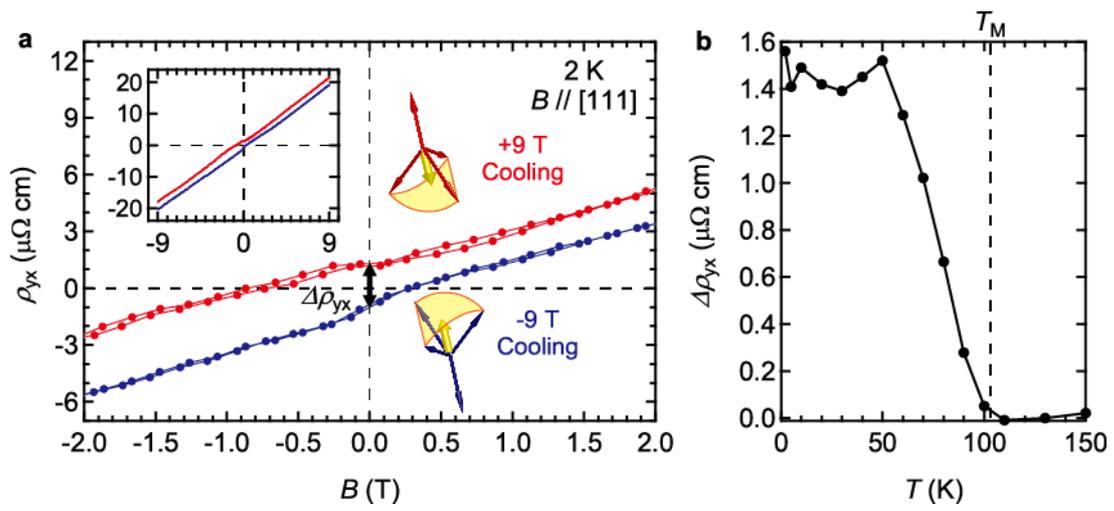

Fig. 4 Fujita et al.,

**Supplementary Materials:**

**Odd-parity magnetoresistance in pyrochlore iridate thin films with broken time-reversal symmetry**

T. C. Fujita[1], Y. Kozuka[1,*], M. Uchida[1], A. Tsukazaki[1,2,3], T. Arima[4,5], M. Kawasaki[1,5].

[1]**Department of Applied Physics and Quantum-Phase Electronics Center (QPEC), University of Tokyo, Tokyo 113-8656, Japan.**

[2]**Institute for Materials Research, Tohoku University, Sendai 980-8577, Japan.**

[3]**PRESTO, Japan Science and Technology Agency (JST), Tokyo 102-0075, Japan.**

[4]**Department of Advanced Materials Science, University of Tokyo, Kashiwa 277-8561, Japan.**

[5]**RIKEN Center for Emergent Matter Science (CEMS), Wako 351-0198, Japan.**

**Growth phase diagram**

Single phase $Eu_2Ir_2O_7$ films were prepared by pulsed laser deposition. Although pyrochlore phase has never been formed in as-grown films, we succeeded in obtaining single crystalline phase by solid state epitaxy as well as optimizing laser energy, composition of the target, and anneal condition. **a**, Substrate temperature and oxygen partial pressure ($P_{O2}$) during pulsed laser deposition were controlled as indicated in the phase diagram shown in top panel of Fig. S1a. The diagram can be divided into 4 regions based on the results of XRD as shown in Figs. S1b – S1e for representative films. **b**, Only $Eu_2O_3$ peaks appeared. Compositional analysis revealed the absence of Ir, possibly due to the evaporation of Ir containing spices, presumably $IrO_3$ with a very high vapor pressure[22]. **c**, In addition to the peaks of $Eu_2O_3$, those of metallic Ir appeared, indicating too reductive conditions. **d**, Both Ir and Eu are in amorphous phase as detected by energy dispersive x-ray spectroscopy. **e**, Single phase $Eu_2Ir_2O_7$ appeared by annealing in an electrical furnace at 1000 °C for 1.5 hours in air. Samples within **d** region but outside of **e** region shown by a box contained some impurity phases such as Ir metal and polycrystalline $Eu_2O_3$ even after annealing under the same conditions. Broken lines indicate the phase coexistence conditions for respective pairs of oxidation states of iridium, which were determined by electrochemical calculation.

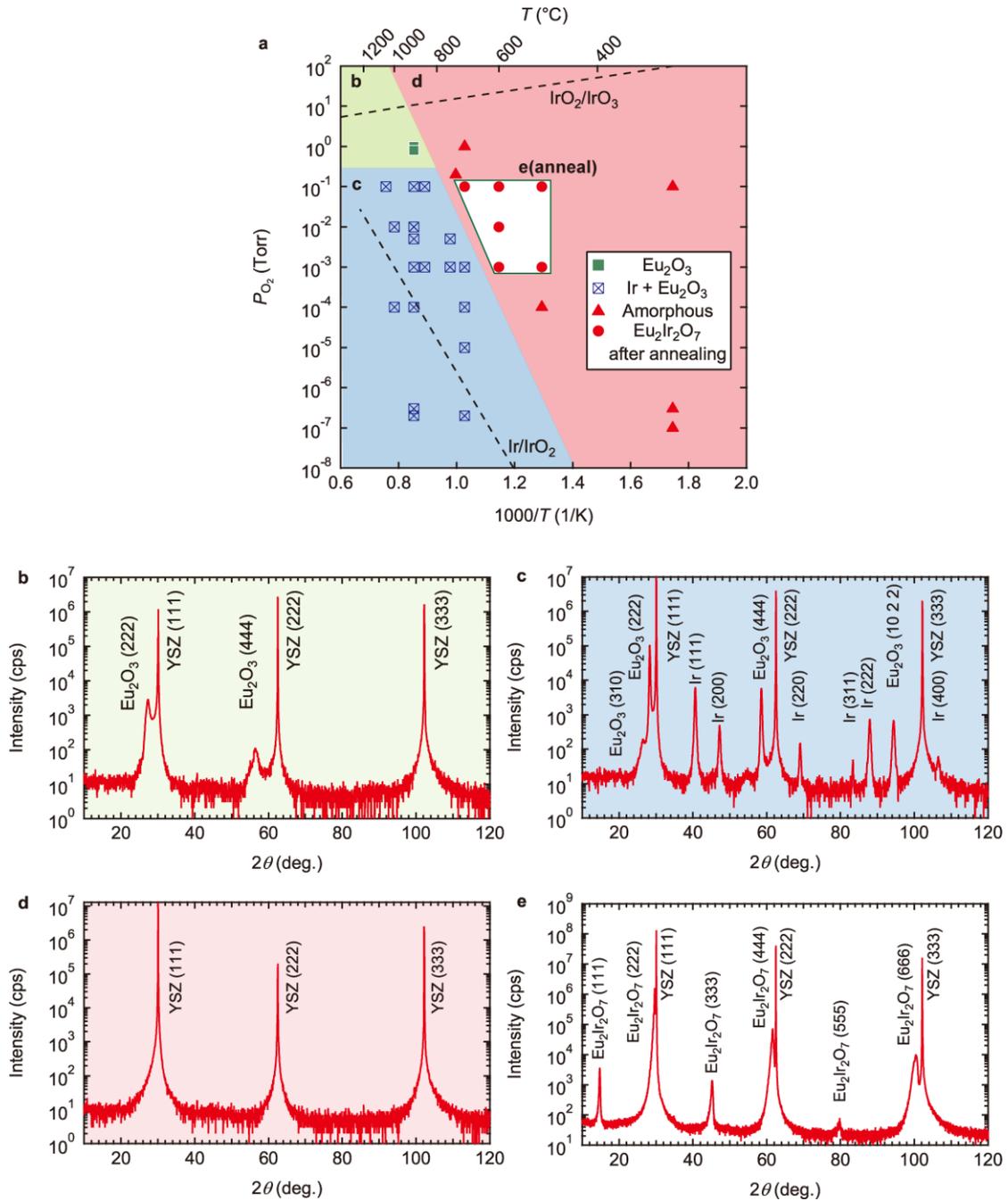

**Figure S1 |A phase diagram of Eu$_2$Ir$_2$O$_7$ films grown on YSZ (111) substrate. a**, The growth phase diagram as functions of $P_{O2}$ and temperature, classified into the following four regions based on x-ray diffraction. **b**, Ir evaporates and only Eu remains in the form of Eu$_2$O$_3$. **c**, Ir is crystallized and Eu forms polycrystalline Eu$_2$O$_3$. **d**, Both Ir and Eu are in amorphous phase. **e**, Eu$_2$Ir$_2$O$_7$ phase appears after annealing although it is amorphous in the as-grown state. Broken lines indicate the phase coexistence conditions for respective pair of iridium related compounds.

**Structural and Electrical characterization:**

Additional XRD results are shown in Fig. S2. The rocking curve around Eu$_2$Ir$_2$O$_7$ (222) peak is shown in Fig. S2a. The full width at half maximum (FWHM) is 0.085°. Azimuthal ($\phi$) scan around Eu$_2$Ir$_2$O$_7$ (662) in Fig. S2b shows three-fold symmetry, which is the same as that of the YSZ substrate, indicating that our films contain neither misoriented domain nor stacking faults. Different peak intensities at different $\phi$ are due to the slight misalignment of the sample.

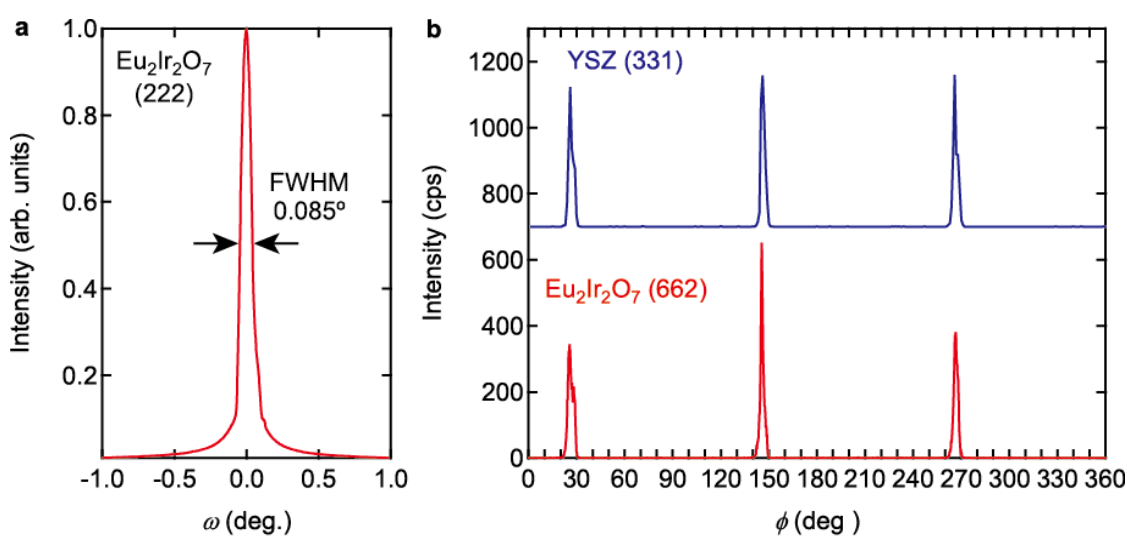

**Figure S2 |Rocking curve and azimuthal scan. a**, Rocking curve around Eu$_2$Ir$_2$O$_7$ (222) peak. **b**, Azimuthal scan around Eu$_2$Ir$_2$O$_7$ (662) peak. The Eu$_2$Ir$_2$O$_7$ film shows three-fold symmetry.

Atomic force microscope (AFM) image of the sample before and after annealing in the region of 2 μm square are shown in Fig. S3. The route mean square roughness (RMS) is order of 1 nm for both before and after annealing. Although surface morphology is quite different between before and after annealing, RMS is almost the same.

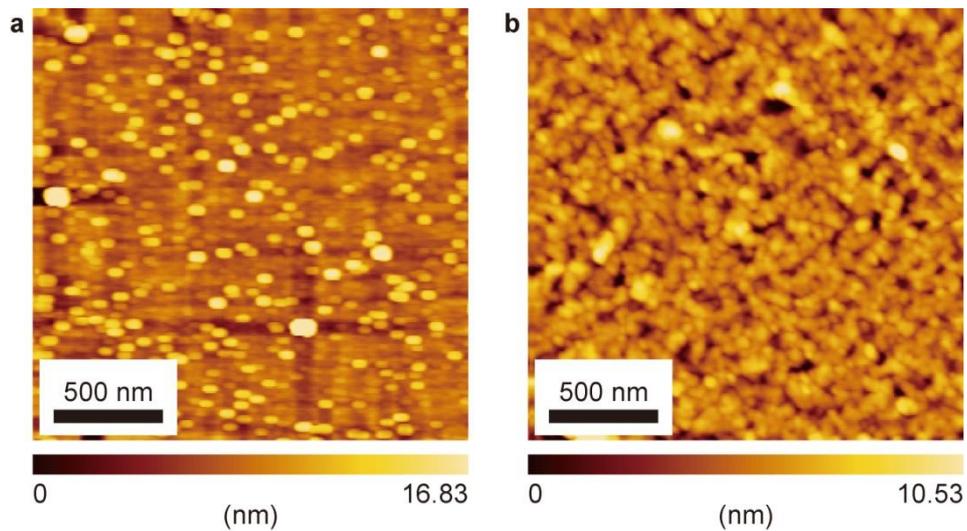

**Figure S3 |AFM image before and after annealing. a**, before and **b**, after annealing.

Figure S4 shows temperature dependence of the longitudinal resistivity ($\rho_{xx}$) and $\rho_{xx}$(2 K)/$\rho_{xx}$(300 K) for typical Eu$_2$Ir$_2$O$_7$ thin films. In the main part of the paper, the data obtained in No. 3 sample are discussed. The large variation of $\rho_{xx}$ and strength in MIT are thought to be originated from the cation off-stoichiometry that was controlled by varying Ir/Eu composition in the target; Ir/Eu = 1 and 1.5 for samples 1 and 2, respectively. Other growth conditions were the same for all the samples. From a simple comparison with bulk polycrystalline samples in the previous report[25], our samples are off-stoichiometric by about 1 – 4 % Ir-rich.

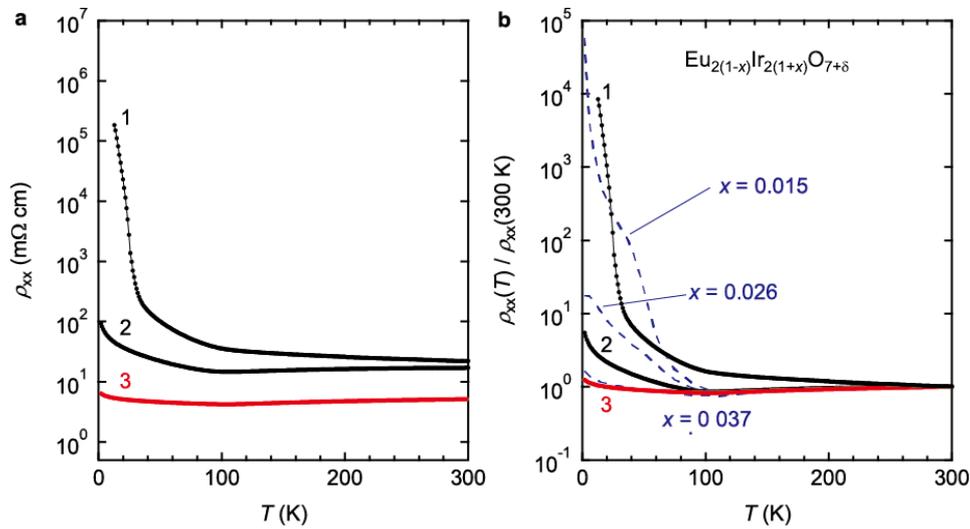

**Figure S4 |Temperature dependence of $\rho_{xx}$ and $\rho_{xx}$(*T*)/$\rho_{xx}$(300 K).** The data for three representative samples are shown. The sample 3 and highlighted by red colour is discussed in the main part. The data for bulk polycrystalline samples in the previous paper[25] are also shown in **b** as dashed lines for comparison. Chemical composition is expressed in the formula Eu$_{2(1-x)}$Ir$_{2(1+x)}$O$_{7+\delta}$, where $\delta$ is thought to be 0 (ref. 25).

Figure S5a displays the symmetric and asymmetric terms in $\rho_{xx}$ extracted from the data in Fig. 3a. The symmetric terms completely coincide for three different cooling conditions, and can be interpreted as ordinary magnetoresistance possibly originating from delocalization by orbital effect. In contrast, asymmetric terms only appear in field cooling data as shown in Fig. S5b. This asymmetric term is dominated by a linear term, the polarity of which is switched with inverting the cooling field direction. (A)symmetric terms are calculated by a conventional (anti-)symmetrization defined as below:

$$\rho_{xx} sym = \frac{\rho_{xx}(B) + \rho_{xx}(-B)}{2}$$
$$\rho_{xx} asym = \frac{\rho_{xx}(B) - \rho_{xx}(-B)}{2}$$
(S1)

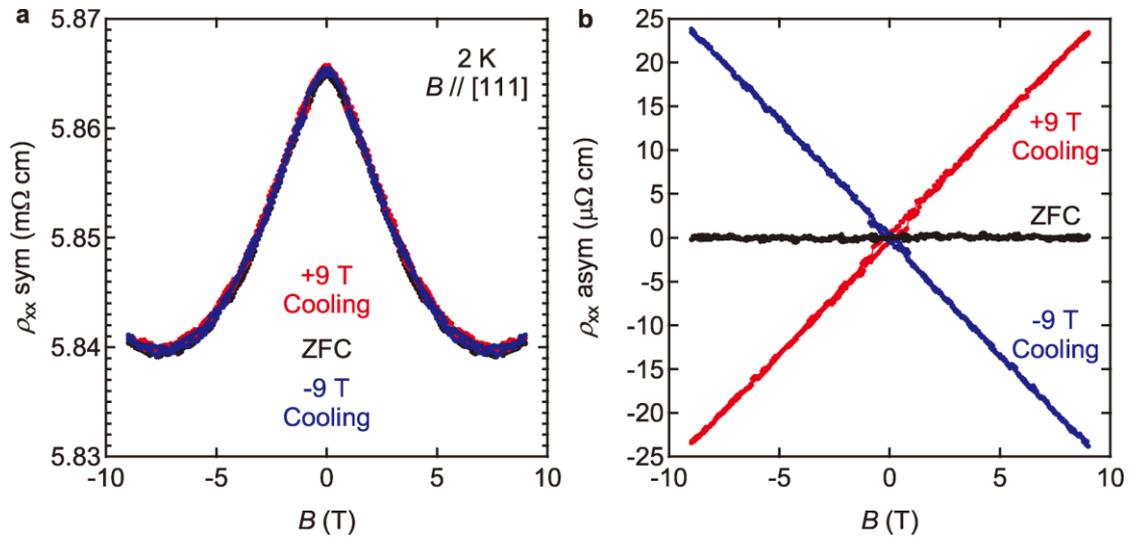

**Figure S5 |Symmetrized and anti-symmetrized magnetoresistance. a**, Symmetric and **b**, asymmetric terms are extracted from the data shown in Fig. 3a.

Figure S6 shows two-terminal longitudinal resistivity ($\rho_{xx2}$) as a function of magnetic field ($B$) to exclude any contribution from $\rho_{yx}$. A linear term is clearly observed in magnetoresistance and the polarity is inverted, depending on cooling field direction. This result rules out the possibility of intermixing with $\rho_{yx}$ as a source of linear term in $\rho_{xx}$.

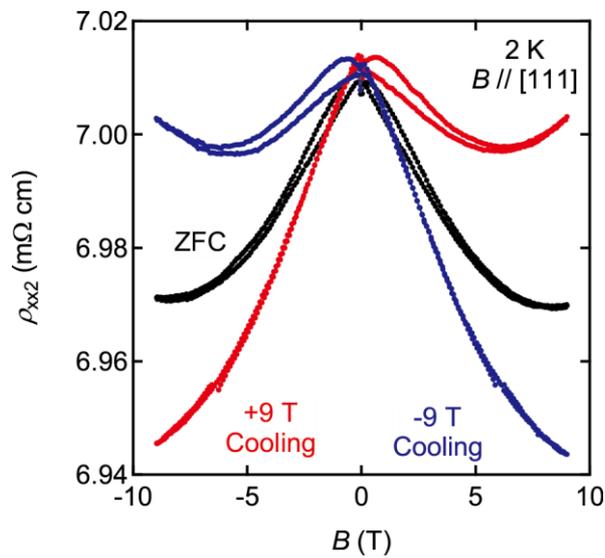

**Figure S6 | Two-terminal resistivity.** Raw data of the two-terminal longitudinal resistivity ($\rho_{xx2}$) at 2 K for strained film after cooling under 0 T (ZFC) and ± 9 T.

In Fig. S7 we show the symmetric (Fig. S7a) and asymmetric (Fig. S7b) term of $\rho_{xx}$ defined by Eq. (S1) for several temperatures under different cooling magnetic fields (0 T and ±9 T). At lower temperatures below 50 K, symmetric term shows almost the same behaviour independent of cooling field and do not have hysteresis. With increasing temperature to 70 K, weak but finite hysteresis is observed, and it disappears again above $T_M$ (105 K). In contrast, the asymmetric term highly depends on cooling field, having exactly opposite sign between +9 T and −9 T, and almost zero for 0 T cooling, and such behaviour is independent of measuring temperature below $T_M$. From these results, we can conclude that (i) the magnetic domain structure is completely fixed at lower temperatures, (ii) above around 60 K, other ferromagnetic ordering may emerge due to the sweep magnetic field, (iii) no magnetic order exists above $T_M$. The coercive field can be extremely high of the order of ~ 100 T in the case of antiferromagnetic frustrated systems, but it is not clear for the present system at the moment.

The temperature of 60 K is consistent with the temperature, at which $\Delta\rho_{yx}$ shown in Fig. 4b is saturated whereas $\alpha$ shown in Fig. 3b is not. One possible origin of this difference is considered to be (i) the development of other magnetic order or (ii) the existence of coupling parameter between carriers and magnetic moments. Particularly for (ii), both carrier conduction and magnetism are originated from $5d$ electron of Ir in this system, which would further enhance such coupling. In order to verify our consideration determining exact magnetic structure for this system is inevitable and this remains as a future work.

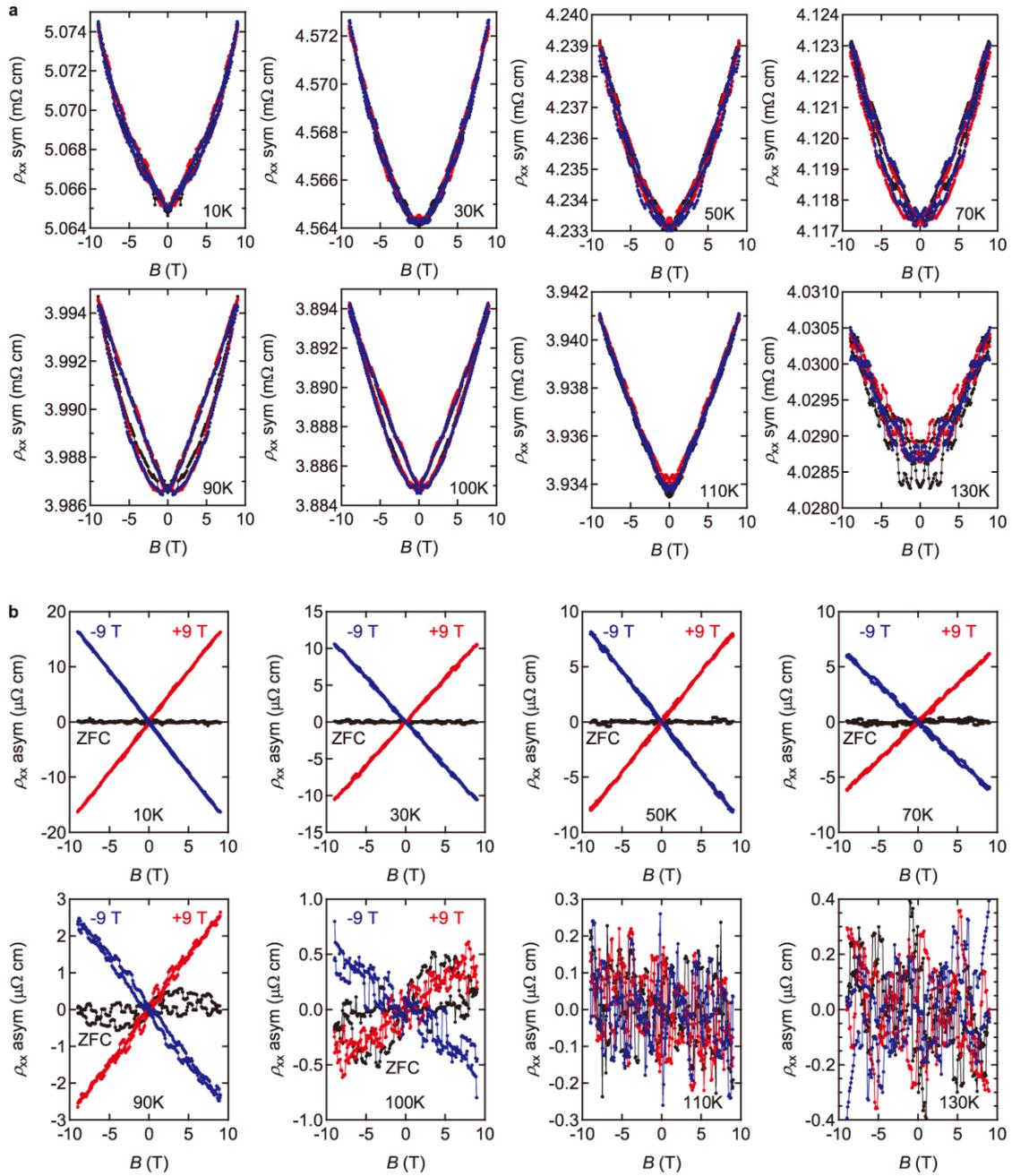

**Figure S7|Temperature dependence of symmetrized and anti-symmetrized magnetoresistance. a**, Symmetric and **b**, asymmetric terms in magnetoresistance shown in Fig. 3a. after cooling under 0 T (ZFC) and ±9 T.

Figure S8a shows the field-asymmetric terms ($\rho_{yx}$ asym) in $\rho_{yx}$ from the data in Fig. 4a, which is calculated by deducing $\Delta\rho_{yx}$. Here, we note that these terms exhibit bending around 0 T, the directions of which are opposite depending on the cooling field, namely magnetic domain structure. By subtracting *B*-linear ordinary Hall term from $\rho_{yx}$, we can extract non-linear components ($\rho_{NL}$) of $\rho_{yx}$ (Fig. S8b). Although the origin is still unclear as mentioned in main text, $\rho_{NL}$ should be intrinsic property in this system reflecting the magnetic structure because the sign change in anomalous Hall effect is not occurred in general by extrinsic sources such as magnetic impurities. One of the possible reasons for these observations is that band crossing near the Fermi-level generates Berry curvature which induces fictitious magnetic flux as reported in the case of $EuTiO_3$ (Ref. 33) or $SrRuO_3$ (Ref. 34), and this would be validated by carrier level control with using field-effect transistor technique for our thin films. Another possibility is the existence of the isolated magnetic domains at the interface/surface of the film as pointed in main text, because inversion symmetry is spontaneously broken there and net magnetic moment could remain. In this case, $\rho_{NL}$ is affected by the termination (kagome plane or triangular plane) and dimensionality (should be enhanced in thinner film) of the film, and further refinements of on growth technique is inevitable.

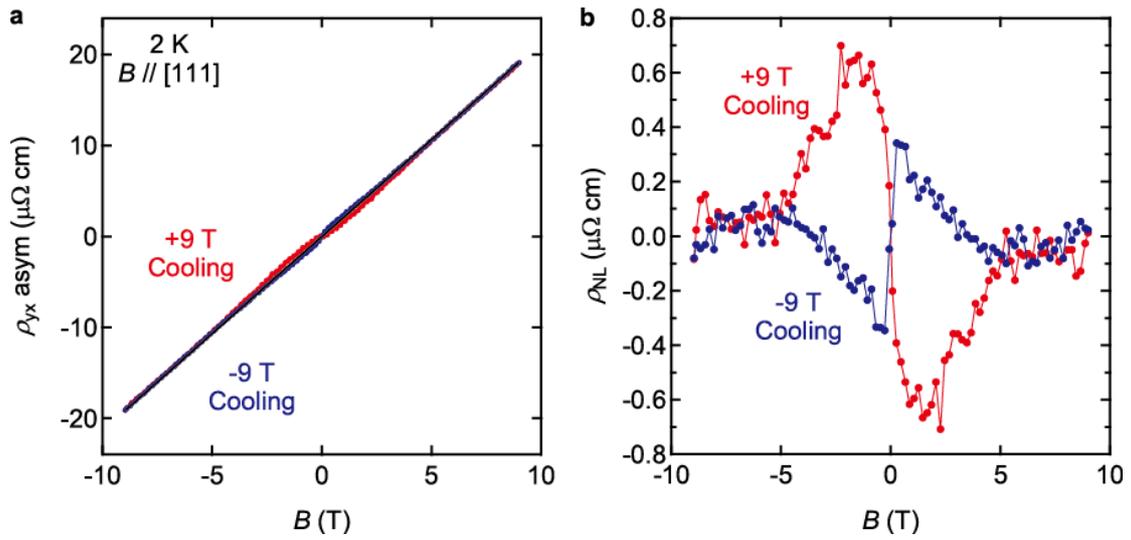

**Figure S8 |Non-linear term in Hall resistivity. a,** Asymmetric and **b,** non-linear components ($\rho_{NL}$) in Hall resistivity at 2 K for strained film after ±9 T of magnetic field cooling from the data in Fig. 4a.